\documentclass[twocolumn,showpacs,prl,floatfix,superscriptaddress]{revtex4}
\usepackage{amsmath,amssymb,eucal}
\usepackage{graphicx}
\usepackage{float}
\usepackage{multirow}
\begin{document}
\title{Zero-Temperature Freezing in 3d Kinetic Ising Model}
\author{J. Olejarz}
\affiliation{Center for Polymer Studies
and Department of Physics, Boston University, Boston, MA 02215, USA}
\author{P. L. Krapivsky}
\affiliation{Center for Polymer Studies
and Department of Physics, Boston University, Boston, MA 02215, USA}
\affiliation{Institut de Physique Th\'eorique CEA, IPhT, F-91191 Gif-sur-Yvette, France}
\author{S. Redner}
\affiliation{Center for Polymer Studies
and Department of Physics, Boston University, Boston, MA 02215, USA}

\begin{abstract}
  We investigate the long-time properties of the Ising-Glauber model on a
  periodic cubic lattice after a quench to zero temperature.  In contrast to
  the conventional picture from phase-ordering kinetics, we find: (i) Domains
  at long time are highly interpenetrating and topologically complex, with
  average genus growing algebraically with system size.  (ii) The long-time
  state is almost never static, but rather contains ``blinker'' spins that
  can flip {\em ad infinitum} with no energy cost.  (iii) The energy
  relaxation has a complex time dependence with multiple characteristic time
  scales, the longest of which grows exponentially with system size.
\end{abstract}
\maketitle

Phase ordering kinetics is concerned with the growth of domains of ordered
phase when a system is suddenly cooled from a high-temperature
spatially-homogeneous phase to a subcritical temperature~\cite{GSS83,B94,book}.
For systems with a non-conserved order parameter, single-phase regions emerge
and form a coarsening domain mosaic whose typical length scale grows in time
as $t^{1/2}$.  This growth continues until the system reaches the equilibrium
state with a non-zero order parameter.  An archetypical example is the Ising
model that is endowed with Glauber dynamics~\cite{G63}, where domains consist
of contiguous regions of spins that all point up or point down.

What happens when the final temperature is zero?  While an infinite system
will coarsen indefinitely, coarsening should stop in a finite system of
linear dimension $L$ when the typical domain length becomes comparable to
$L$.  A natural expectation is that the ground state will ultimately be
reached, and indeed this outcome occurs in the one-dimensional Ising-Glauber
model~\cite{B94,book}.  Surprisingly, this conventional picture already
begins to fail in two dimensions where the ground state is reached
roughly two-thirds of the time; in the remaining cases, the system falls into
an infinitely long-lived metastable state that consists of two (and more than
two in rare cases) straight single-phase stripes~\cite{SKR01,ONSS06,BKR09}.

The fate of the three-dimensional Ising ferromagnet with zero-temperature
Glauber dynamics is even more intriguing.  First, the long-time state is
topologically complex, with multiply-connected interpenetrating regions of
positive and negative magnetization.  This sponge-like geometry represents a
discrete analog of zero average-curvature interfaces, for which a veritable
zoo of possibilities have been cataloged~\cite{S90,S69}.  There is also a
close resemblance to gyroid phases, or ``plumber's nightmares'' that arise in
micellar and other two-phase systems~\cite{L89,AS89,FUTGW,AECTB}.  Second,
even though the temperature is zero, almost all realizations fluctuate
forever due to {\em blinker spins} --- a subset of spins that can flip
repeatedly without any energy cost~\cite{SKR01}.  Last, the approach to these
asymptotic blinker states is extraordinarily slow, with relaxation times that
grow exponentially with system size.  If the initial magnetization is
non-zero, it is believed that the ground state of the initial majority phase
is reached~\cite{SKR01,M10}, while if the boundary spins have a fixed sign
the ground state is always reached~\cite{CMST}.

Our system is the three-dimensional homogeneous ferromagnetic Ising model on
a cubic lattice of linear dimension $L$ with periodic boundary conditions.
The system is initialized in the antiferromagnetic state~\cite{AF,OKR10} and
spins subsequently evolve by zero-temperature Glauber dynamics: a
randomly-selected spin flips with probability 1 if the energy of the system
decreases, flips with probability $\frac{1}{2}$ if the energy does not
change, and does not flip if the energy were to increase.

\begin{figure}[ht]
\includegraphics[width=0.25\textwidth]{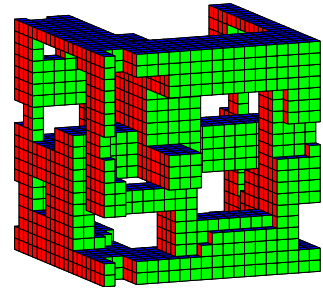}
\caption{(color online) Example of a genus $g=17$ final-state domain on a
  $20^3$ lattice with periodic boundary conditions.  Each block represents an
  up spin (with the spin at the center of the block), while blank space
  represents a down spin.}
\label{g-20}
\end{figure}

\noindent{\tt Energy and Topological Complexity:} A fundamental
characteristic of the long-time state is the dependence of the energy gap
(per spin) $E_L$ versus system size $L$.  Even though the system does not
reach the ground state, the energy systematically decreases with $L$.  Direct
simulations to reach the asymptotic state of even medium-size systems are
prohibitively slow, however, because energy-lowering spin-flip events become
progressively more rare once the coarsening length scale reaches the system
size.  In this post-coarsening regime, the energy evolution is characterized
by long periods where only zero-energy spins (those with equal numbers of up
and down neighbors) flip, punctuated by rare energy-decreasing events.

To reduce the time needed to simulate these long iso-energy wanderings, we
employ an acceleration protocol: Once energy-lowering events
become rare, we apply an infinitesimal magnetic field as the system wanders
on each energy plateau between energy-lowering events~\cite{field}.  The 
field drives the state-space motion on each plateau so that the
next energy-lowering spin flip is found more quickly.  After each
energy-lowering event, the direction of the infinitesimal field is reversed
so that the net average field is zero.  We verified that this procedure
accurately reproduces the energy that is obtained by Glauber dynamics for the
range of system sizes ($L\leq 10$) where a direct check of this acceleration
method is computationally feasible~\cite{OKR10}.  (The relative deviation of
the average energies obtained by these two approaches is less than $10^{-7}$ for size
$10^3$, while taking a factor of 58 less CPU time.)

Our data for the energy is based on systems of linear dimension $L\leq 76$
with $\geq 10^5$ realizations for each value of $L$.  The relative error for
each data point is $< 0.1\%$.  Our data are consistent with $E_L\sim
L^{-\epsilon}$ with $\epsilon\approx 1$, in agreement with previous results
based on smaller-scale simulations~\cite{SKR01}.  This dependence implies
that the total interface area between spin domains scales as $L^2$.

At long times, there are almost always just two interpenetrating
domains~\cite{OKR10}, and these domains are topologically complex.  We
quantify a domain topology by its genus $g$, which equals the number of holes
in the domain surface.  (The genus of a sphere is $g=0$, while the genus of a
doughnut is $g=1$.)~ Figure \ref{g-20} shows a large-genus example (with
$g=17$) for a $20^3$ periodic system.  To measure the genus of a domain, we
exploit the connection to the Euler characteristic~\cite{E}
\begin{equation}
\label{Euler}
\chi= 2(1-g)= V-E+F
\end{equation}
that relates $\chi$ to easily-measured features of the interface: $V$, the
number of vertices on the interface, $E$, the number of edges, and $F$, the
number of faces.  Each face separates a pair of oppositely-oriented
neighboring spins, so that $F$ is directly related to the energy by $F\sim
L^3 E_L$.  Our simulation data for systems with $L\leq 76$ again show
considerable finite-size corrections but extrapolation suggests that the
average genus grows as $\langle g\rangle\sim L^\gamma$ with $\gamma\approx
1.7$.

The final energy $E_L$ leads to an upper bound on the genus $g$.  To
establish this bound, we simplify Eq.~\eqref{Euler} by noting that a face has
4 edges, and each edge is shared between 2 adjacent faces.  Hence $E=2F$.
Similarly, each edge has 2 vertices that are shared among 3, 4, or 5 adjacent
edges, giving $\frac{2}{5}E\leq V\leq \frac{2}{3}E$.  Using these relations
in Eq.~\eqref{Euler} gives $-\frac{1}{5}F\leq\chi\leq \frac{1}{3}F$, or
$0\leq g\leq 1+\frac{1}{10}F$ where we additionally use that the number of
holes $g$ is non-negative.  From the relation between the number of faces and
the energy, $F\sim L^3 E_L\sim L^{3-\epsilon}$, with $\epsilon\approx 1$, we
have the upper bound $g\leq L^{3-\epsilon}$.  This growth rate is slightly
faster than that suggested by our simulations, $\langle g\rangle\sim
L^\gamma$, with $\gamma\approx 1.7$.  \smallskip

\begin{figure}[ht]
\begin{center}
\includegraphics[width=0.25\textwidth]{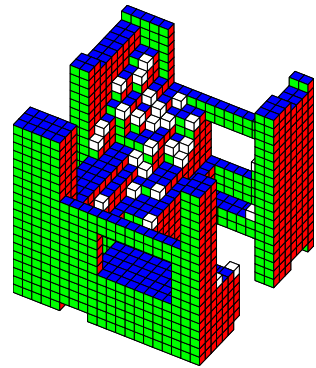}
\caption{\small (color online) An example of a blinker state on a $20^3$
  cubic lattice with periodic boundary conditions.  Highlighted blocks
  correspond to ``blinker'' spins that point up.}
\label{sponge}
  \end{center}
\end{figure}

\noindent{\tt Blinker States:} As the linear dimension is increased, it
becomes overwhelmingly likely that the system does not reach a static state
at long times, but rather, gets trapped within a set of perpetually evolving
configurations that contain stochastic {\em blinker} spins.  In
Fig.~\ref{sponge}, spins that point up lie at the center of the small cubes
and blinker spins are highlighted, while down spins are represented by blank
space.  Equivalently, spin-up blinkers are located at the convex (outer)
corners of the interface, while spin-down blinkers are adjacent to the apex
of the concave (inner) corners.  Each blinker spin has three neighboring
spins of the same sign and three of the opposite sign so that a blinker can
flip without changing the energy of the system.  When a blinker spin flips,
one of its neighbors typically becomes a blinker so that blinkers never cease
to evolve.

A system that contains blinker spins can therefore wander forever on a small
set of iso-energy points in state space.  We define this set as a {\em
  blinker state}.  While the fraction of blinker spins is small --- typically
less than a percent when the linear dimension $L\geq 10$ --- the fraction of
the system volume over which blinker spins can wander is roughly 9\% for
large $L$~\cite{OKR10}.  

These blinkers are part of the huge number of metastable states in the
system.  For example, the number of metastable states that consist of
alternating slabs of plus and minus spins (the three-dimensional analog of
stripe states in two dimensions) scales as $\exp(L^2)$~\cite{SKR01}; a
similar estimate for the number of metastable sponge-like states gives
$\exp(L^3)$~\cite{AP}.  Thus it is plausible that the three-dimensional Ising
model with Glauber dynamics should get trapped in one of these ubiquitous
metastable states.  What is striking, is that the system almost always falls
into a perpetually evolving blinker state, rather than a static metastable
state.  For example, for $10^5$ realizations at $L=76$, the fraction of
realizations that end in a blinker state, a static metastable state, and the
ground state are 97.46\%, 2.50\%, and .04\%, respectively.  \smallskip

\begin{figure}[ht]
\begin{center}
\includegraphics[width=0.15\textwidth]{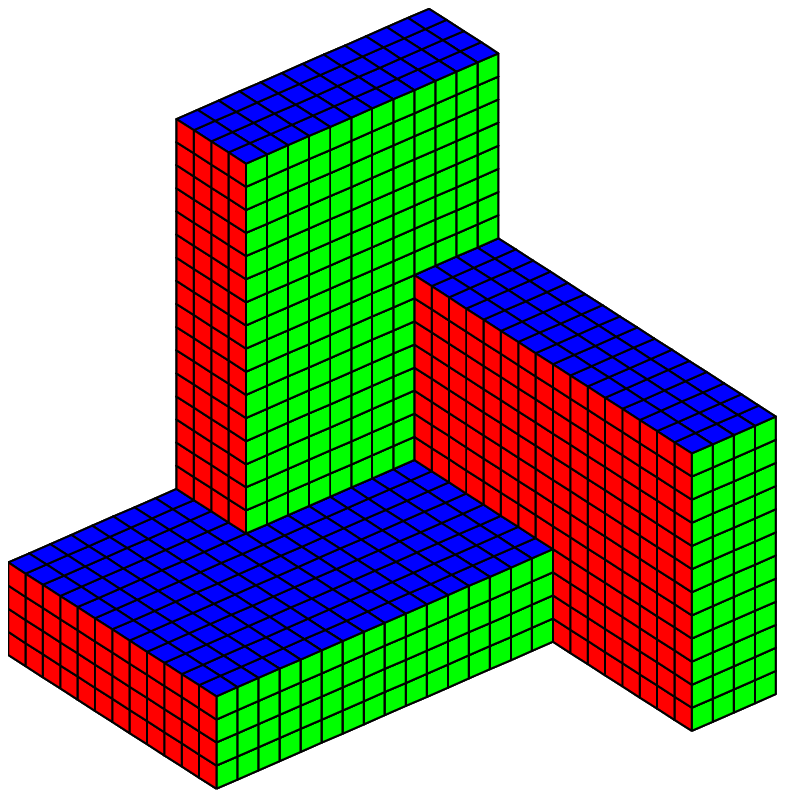}\hskip 0.2cm
\includegraphics[width=0.15\textwidth]{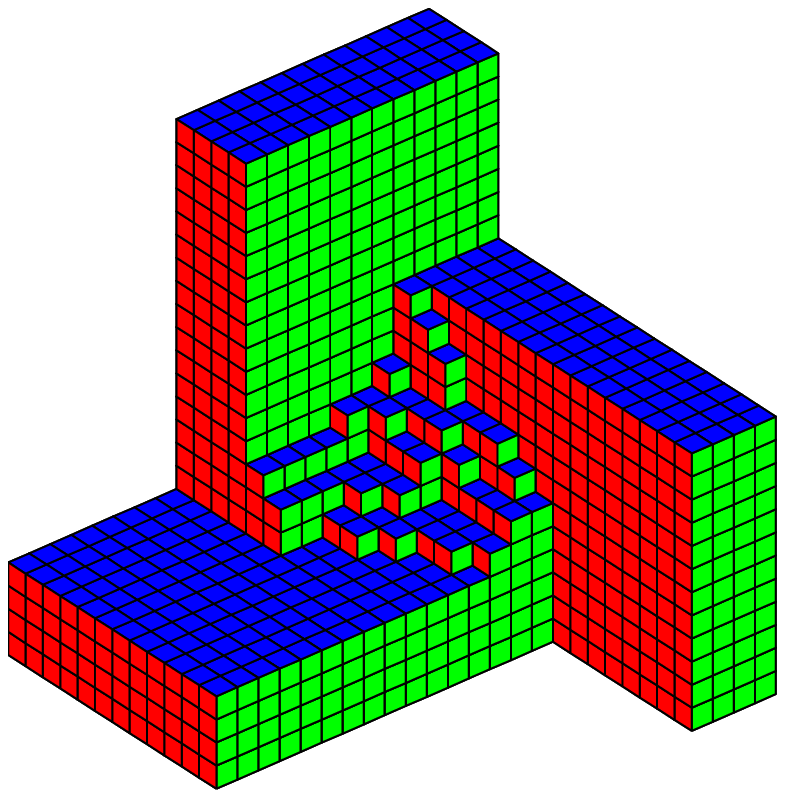}\hskip 0.2cm
\includegraphics[width=0.15\textwidth]{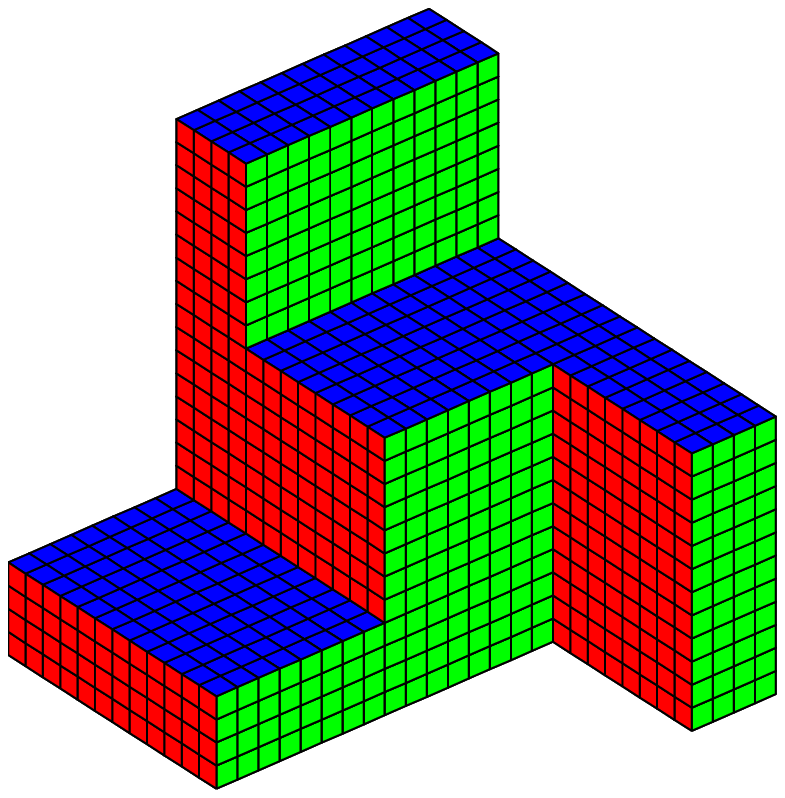}
\smallskip
\caption{\small (color online) An $8\times 8\times 8$ blinker on a $20^3$
  cubic lattice, showing the fully-deflated (left), intermediate (middle),
  and fully-inflated states (right).  The bounding slabs wrap periodically in
  all three Cartesian directions.}
\label{blinker-evol}
  \end{center}
\end{figure}

\noindent{\tt Ultra-Slow Relaxation:} Blinker states are also responsible for
an extremely slow relaxation that involves time scales that grow faster than
power law in the system size~\cite{slow}.  To understand the cause of these
long time scales, consider the synthetic cubic blinker state shown in
Fig.~\ref{blinker-evol}.  By zero-energy spin flips, the interface defined by
the blinker spins can be in the extremes of fully deflated (left),
intermediate (middle), or fully inflated (right).  Although each blinker spin
does not have any energetic bias, there exists an effective geometric bias
that drives the interface to the half-inflated state.  This effective bias
stems from the difference in the number of flippable spins on the convex
(outer) and concave (inner) corners on the interface, $N_+$ and $N_-$,
respectively.  When the interface is mostly inflated, $N_+-N_-$ is positive,
so that there are, on average, more spin flip events that tend to deflate the
interface, and vice versa when the interface is mostly deflated.  This
effective bias drives the interface to the half-inflated state.

\begin{figure}[ht]
\begin{center}
\includegraphics[width=0.11\textwidth]{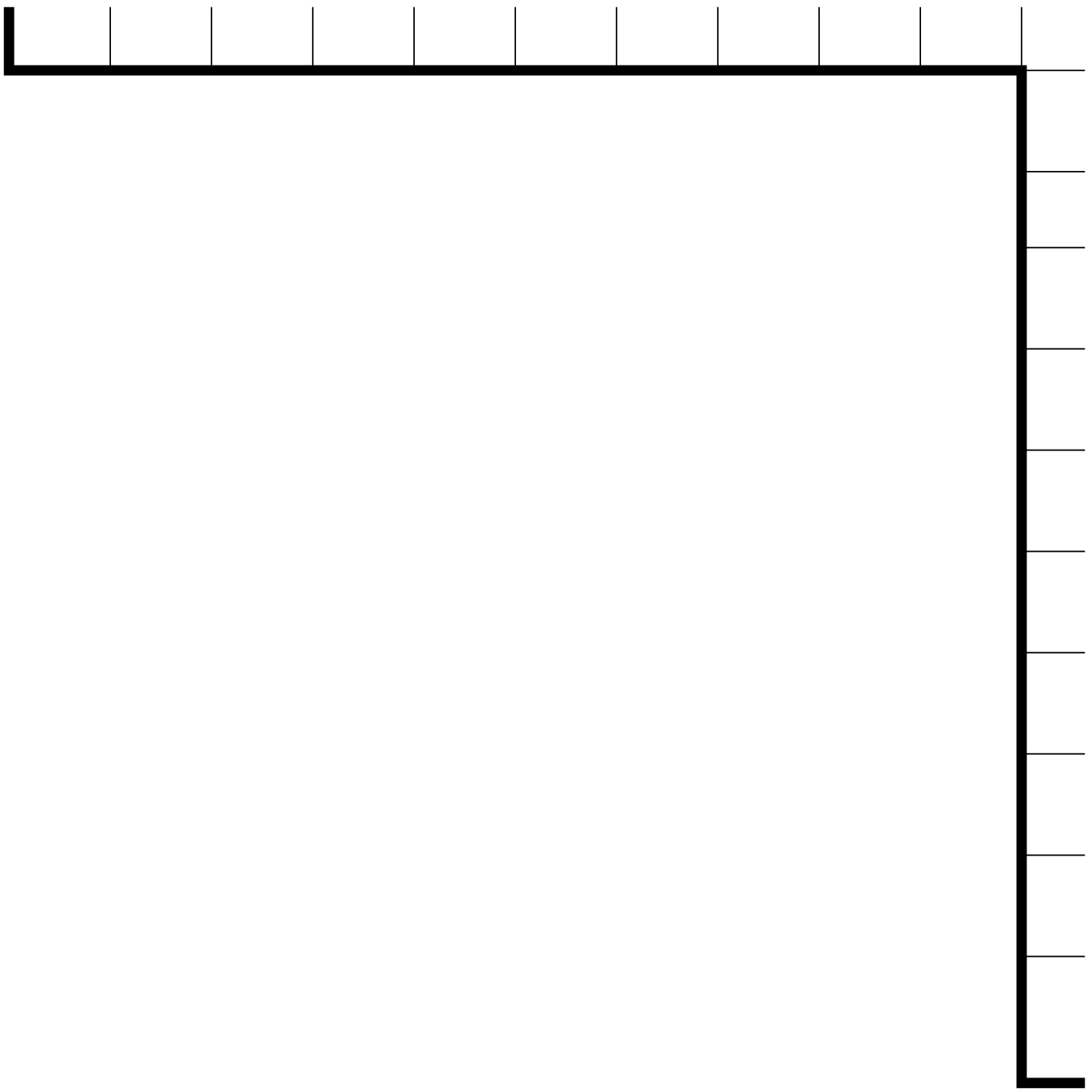}\quad\qquad
\includegraphics[width=0.11\textwidth]{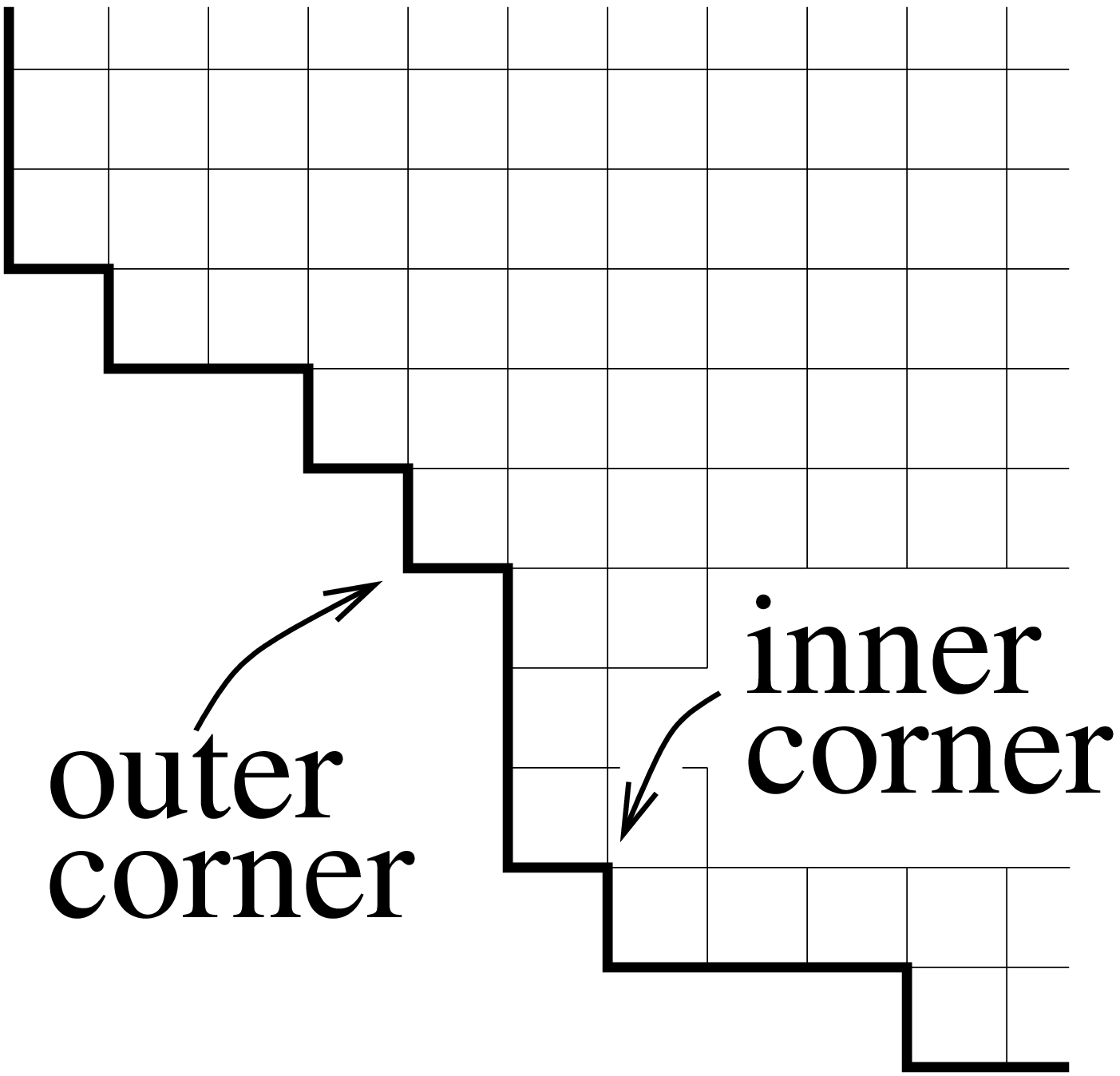}\quad\qquad
\includegraphics[width=0.11\textwidth]{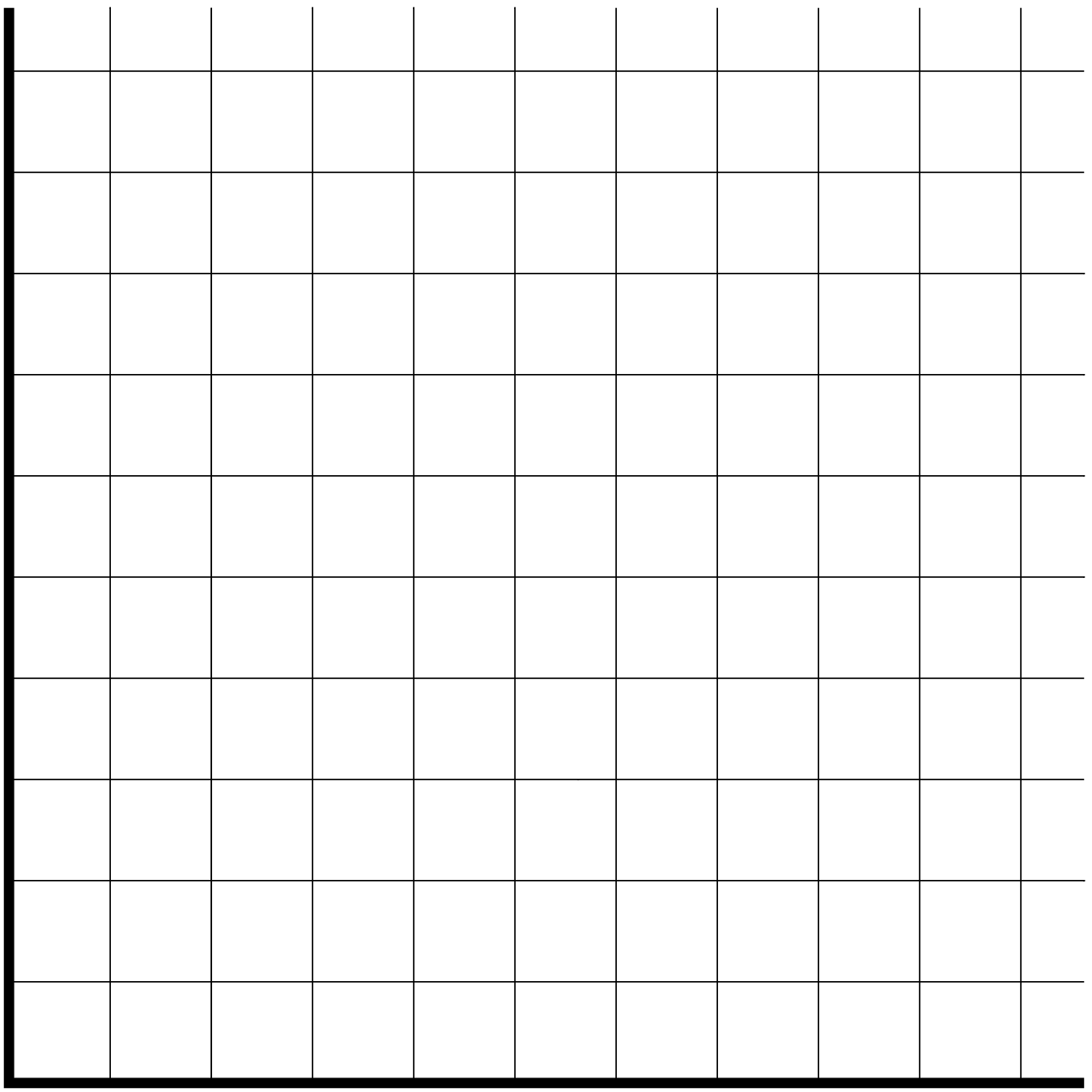}
\smallskip
\caption{\small Two-dimensional analog of the blinker states in
  Fig.~\ref{blinker-evol}. }
\label{blinker-2d}
  \end{center}
\end{figure}

We quantify the relaxation of this blinker by the first-passage time $\langle
t\rangle$ for an $\ell\times \ell\times \ell$ half-inflated blinker
(Fig.~\ref{blinker-evol} middle) to reach the fully-inflated state.  For
simplicity, consider first the corresponding two-dimensional system
(Fig.~\ref{blinker-2d}).  Near the half-inflated state, the interface
consists of $N_+$ outer corners and $N_-$ inner corners, with $N_+-N_-$
always equal to 1 in two dimensions, and $N_+\sim \ell$~\cite{TKR04}.  In one
time unit, all eligible spins on the interface flip once, on average.  Since
$N_+-N_-=1$, the area occupied by the up spins typically decreases by 1.
Thus we infer an interface velocity $u=\Delta A/\Delta t\sim -1$.  Similarly,
since there are $N_++N_-\sim N_+$ spin flip events in one time unit, the
mean-square change in the interface area is of the order of $N_+\sim
\sqrt{A}\sim\ell$.  Thus the effective diffusion coefficient is $D\sim \ell$.
The underlying first-passage process from the half-inflated to the
fully-inflated state requires moving against the effective bias velocity by
flipping $\ell^2/2$ spins to point up.  Consequently, the dominant Arrhenius
factor in the first-passage time is $\tau\sim \exp(|u| \ell^2/2D)$, so
that~\cite{fpp}
\begin{equation}
\label{T2}
\ln \tau\sim  \ell\,.
\end{equation}

For the corresponding three-dimensional blinker, the inflated region is a
cube of volume $\ell^3$.  There are typically $N_\pm\sim \ell^2$ outer 
and inner corners on the interface when it is half
inflated.  In distinction with two dimensions, there is no conservation law
for the difference $N_+-N_-$.  Rather, the disparity between $N_+$ and $N_-$
is of the order of $\ell$.  If the system is beyond the
half-inflated state, then in a single time step the interface will recede, on
average, by $\ell$, giving an interface velocity $u\sim \ell$.  Similarly we estimate
$D\sim N_\pm\sim \ell^2$, leading to
\begin{equation}
\label{T3}
\ln \tau\sim u \ell^3/D\sim \ell^2\,.
\end{equation}
The straightforward generalization to $d$ dimensions gives $\ln \tau\sim
\ell^{d-1}$.  Our simulations~\cite{OKR10} for this first-passage time in
two dimensions agree with Eq.~\eqref{T2}.  In three dimensions, simulations
are necessarily limited to small $\ell$, while our crude argument is
asymptotic.  Moreover, the bias velocity in three dimensions is not strictly
constant during the inflation of the interface, while the bias is constant in
two dimensions.  Nevertheless, the meager data that we do have (up to
$\ell=5$) are qualitatively consistent with Eq.~\eqref{T3}.  The salient
result is that the time for a half-inflated blinker to reach the
fully-inflated state grows extremely rapidly with $\ell$.

\begin{figure}[ht]
\includegraphics[width=0.325\textwidth]{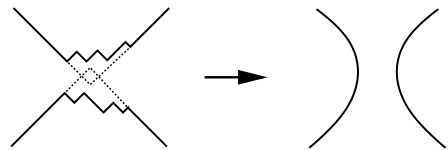}
\caption{Two-dimensional sketch of blinker coalescence. }
\label{blinker-merge}
\end{figure}

From the dynamics of the cubic blinker of Fig.~\ref{blinker-evol}, we can
understand the long-time relaxation of a large system.  Indeed, suppose that
there are two such blinkers that are oppositely oriented and spatially
separated so that they do not overlap when both are half inflated, but just
touch corner to corner when both are inflated (Fig.~\ref{blinker-merge}).  As
long as the blinkers do not overlap, their fluctuations do not change the
energy of the system.  However, if these blinkers touch, then a spin flip
event has occurred that lowers the energy.  After this irreversible
coalescence event, subsequent spin flips cause the two blinkers to ultimately
merge.  Each of these initial coalescences corresponds to one of the
increasingly rare energy-lowering spin-flip events at long times.

\begin{figure}[ht]
\includegraphics[width=0.375\textwidth]{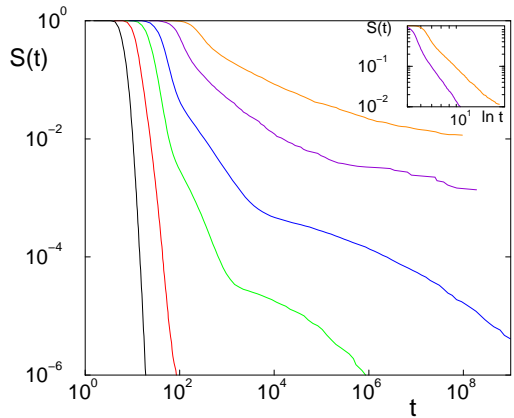}
\caption{(color online) Survival probability for $L=4,6,8,10, 14$ and 20
  (lower left to upper right) for $10^7$ realizations $L\leq 10$ and 10240
  realization for $L=14$ and 20 on a double logarithmic scale.  The inset
  shows $S(t)$ versus $\ln t$ for $L=14$ and 20 on a double logarithmic
  scale.}
\label{S-small}
\end{figure}

To quantify this long-time relaxation, we study $S(t)$, the probability that
the energy of the system is still decreasing at time $t$
(Fig.~\ref{S-small}).  In two dimensions, this survival probability decays as
$S\sim e^{-t/\tau}$, with a decay time $\tau(L)$ that suddenly changes from a
quadratic dependence on $L$ to a faster dependence after $S(t)$ has decayed
by roughly two orders of magnitude~\cite{SKR01,PRL87}.  This transition is
caused by long-lived but finite-lifetime metastable configurations in which
the spins organize into two diagonal stripe domains that wind around the
torus once in both the toroidal and poloidal axes~\cite{SKR01}.

The corresponding behavior in three dimensions is more enigmatic and suggests
multiple temporal regimes.  At short times, $S(t)$ is nearly constant because
the coarsening length scale has not yet reached $L$.  Subsequently $S(t)$
first decays rapidly with time, and eventually more slowly.  As shown in
Fig.~\ref{S-small}, nearly 10\% of all realizations of a $20^3$ system are
still evolving at $t=10^4$ and more than 1\% are still evolving at
$t=10^8$, whereas the coarsening time scale is only 400.  For the largest
systems for which we could simulate $S(t)$ to long times, the decay $S(t)$
seems to reasonably fit an inverse logarithmic law $S(t)\sim (\ln t)^{-1}$
(Fig.~\ref{S-small} inset), a dependence that occurs in simple~\cite{slow}
and glassy spin systems~\cite{RS03}, as well as in granular
compaction~\cite{NKBJN98}.

Thus a basic statistical-mechanics model, the three-dimensional Ising model
with zero-temperature Glauber dynamics, does not reach the ground state.  The
relaxation is extraordinarily slow and almost all realizations eventually
reach a blinker state, which is a set of connected iso-energy points in the
space of metastable states, where the system wanders forever.  Domain
interfaces are topologically complex, with hints of a simple relation between
the genus of domains and the long-time energy.

%We believe that that the same feature occurs in other spin
% systems and even more broadly in situations where the state space has
% sufficient complexity that gradient descent does not find global minima of
% the cost function.  For the Ising-Glauber system, the search for the global
% minimum wanders forever within a set of equal-cost blinker states, leading to
% extraordinarily slow energy relaxation.

\smallskip

We gratefully acknowledge financial support from NSF Grants No.\ DMR-0906504
(JO and SR). 
% and No.\ CCF-0829541 (PLK).


\begin{thebibliography}{99}
\bibitem{AF} Similar results are obtained for equal densities of up and down
  spins (supercritical initial temperature).
\bibitem{field} We begin imposing the field after a time of $5\times L^2$.
\end{thebibliography}
\end{document}